\documentclass[twocolumn,twoside,slac_two]{revtex4}
\usepackage{graphicx}
\usepackage{fancyhdr}
\usepackage{amsmath}
\usepackage{isotope}
\begin{document}

%Title of paper
\title{Solar Neutrino Measurements at Super-Kamiokande-III}

% Repeat the \author .. \affiliation  etc. as needed
%
% \affiliation command applies to all authors since the last
% \affiliation command. The \affiliation command should follow the
% other information

\author{B.S. Yang for the Super-Kamiokande Collaboration}
\affiliation{Department of Physics and Astronomy, Seoul National University, Seoul 151-742, Korea}

\begin{abstract}
The full Super-Kamiokande-III data-taking period, which ran from August of 2006 through August of 2008, yielded 298 live days
worth of solar neutrino data with a lower total energy threshold of 4.5 MeV. During this period we made many improvements to
the experiment's hardware and software, with particular emphasis on its water purification system and Monte Carlo simulations. As
a result of these efforts, we have significantly reduced the low energy backgrounds as compared to earlier periods of detector
operation, cut the systematic errors by nearly a factor of two, and achieved a 4.5 MeV energy threshold for the solar neutrino
analysis. In this presentation, I will present the preliminary SK-III solar neutrino measurement results.
\end{abstract}

%\maketitle must follow title, authors, abstract
\maketitle

\thispagestyle{fancy}

% body of paper here - Use proper section commands
% References should be done using the \cite, \ref, and \label commands
% Put \label in argument of \section for cross-referencing
%\section{\label{}}

%%%%%%%%%%%%%%%%%%%%%%%%%%%%%%%%%%
\section{Introduction\label{secintro}}
\begin{figure}[b]
\centering
\includegraphics[width=80mm]{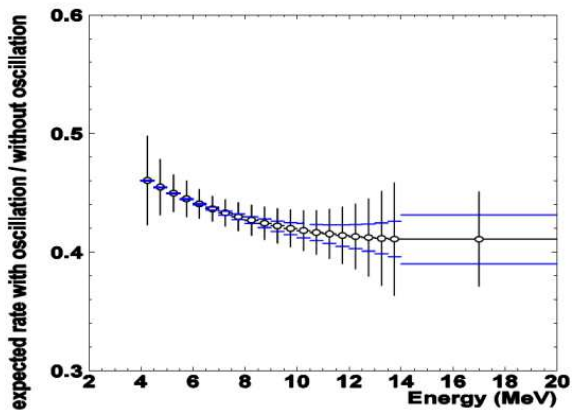}
\caption{Expected spectrum distortion of 5 years SK data with around 70\% reduced background below 5.5 MeV, 4 MeV recoil electron energy threshold and half energy-correlated systematic uncertainty compared to SK-I under current best solution}
\label{fig4}
\end{figure}
 It is well-established that the best solution of the solar neutrino oscillation is MSW-LMA; this is strongly supported by many experiments' results\cite{SKI_full}\cite{SKII_full}\cite{SNONCD}\cite{KAMLAND}\cite{BOREXINO}.  For this solution it is expected that the $\nu_e$ survival probability in the low energy region ($<\sim$1 MeV) is higher than in the high energy region ($>\sim$10 MeV), because for low energies vacuum oscillation is dominant while for high energies matter oscillation is dominant. So a transition from vacuum to matter oscillation should exist, roughly between 1.0 and 10.0 MeV, in the energy range in which \isotope[8]{B} solar neutrinos predominate. The recoil electron energy threshold of Super-Kamioakande is around 5 MeV, so Super-Kamiokande has a chance to discover the low-energy upturn in the \isotope[8]{B} spectrum.

 To realize this goal, reducing background and systematic uncertainty below 5.5 MeV and lowering the recoil electron energy threshold are important \cite{nuclb143.13}. If we can take 5 years of 4 MeV threshold data with 70\% reduced background below 5.5 MeV and half the energy-correlated systematic uncertainty compared to SK-I, we should observe a spectrum as shown in Fig. \ref{fig4} and might discover this energy spectrum distortion.  So, starting with the data of SK-III, making these detector and analysis improvements has been high priority work.

\section{Improvement\label{seccur}}
 Since the end of SK-I, we have worked hard to reduce background and systematic uncertainties, improving several key items. The following sections will introduce some of them.

\subsection{Water System\label{subsecWS}}
\begin{figure}[b]
\centering
\includegraphics[width=80mm]{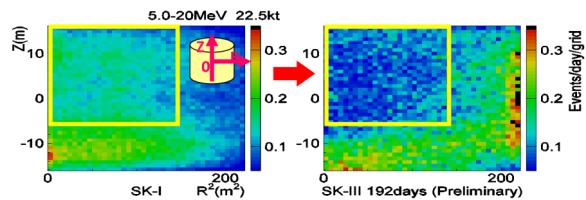}
\caption{Vertex distribution of SK-I/III.}
\label{figver}
\end{figure}
 A major background for the solar neutrino observation at SK is the radioactivity from radon (Rn) in our otherwise extremely pure water.  This ultra-pure water which fills the detector is made from  natural mine water using a very high efficiency water purification system. The Rn background events are very similar to solar events, so it is very difficult to remove them using only analysis tools.  As a result, many Rn events could be included in the final data set. To reduce this background, the improvement of the Rn reduction efficiency of the SK water purification system is the most effective approach. So, since the end of SK-I we have upgraded the system several times, including the addition of a new heat exchanger and two reverse osmosis units for SK-III.
 
 In addition, we investigated the water flow in the tank by purposefully injecting radon-enriched water. Tracing the resulting background events in time from this injected Rn, we found stagnation of water in the top and bottom of the detector volume which increased the background.  To counter this effect we installed new pipes and changed the water flow. Previously, the water was supplied from the bottom of the inner detector (ID) and drained from the top of both the ID and outer detector (OD).  Now, it is supplied from the ID bottom and drained from the top and bottom in the ID and OD, with a total flow two times faster than before. As a result we have a central region with half the  background (yellow box in Fig. \ref{figver}) as compared with SK-I, making lowering of the energy threshold a possibility. 
 
 Note that the excessive background near the wall and bottom of SK-III also existed in SK-II. This background is from the Fiber Reinforced Plastic (FRP) PMT covers which were added at the start of  SK-II in order to protect against propagating shock waves from PMT implosions, and so could not be reduced by improving the water system.  It posed a significant obstacle to enlarging the fiducial volume below 6.5 MeV.

\subsection{Vertex Shift\label{subsecvs}}                                       
\begin{figure}[t]                                                               
\centering                                                                      
\includegraphics[width=80mm]{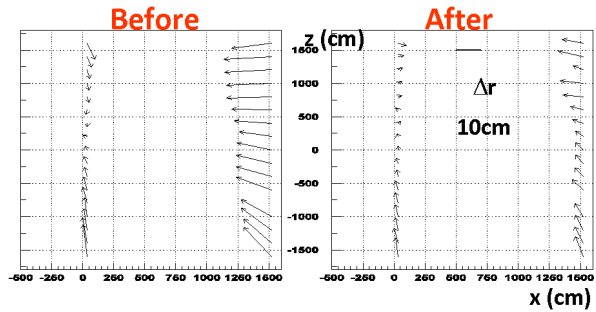}                                        
\caption{Vertex shift before and after correction. After correction,            
vertex shift around wall was reduced to below 10 cm.} \label{figversh}            
\end{figure}                                                                    
The vertex shift, defined as a vector from an averaged position of the reconstructed vertices of the data to that of corresponding Monte Carlo (MC) data, has been one of the main systematic uncertainties for the solar neutrino flux measurement since the beginning of SK-I\cite{SKI_full}. Because it could make events move in or out of the fiducial volume, it is a non-negligible systematic uncertainty.                                                                    

Vertex shift is measured by placing a Ni-Cf gamma ray source at several positions of the tank. The reconstructed data vertices were shifted more than 10 cm from the real source position inward toward the tank center, in contrast those of MC which were shifted less than 10 cm.  The origin of the excessive shift of the data was a mystery since SK-I. In SK-III, we investigated this mystery, resolving it just ten days before the end of SK-III.                   

It turned out that relative hit timing within a wide range ($\sim$100 nsec) was not perfect due to characteristics of our electronics. We measured the timing linearity by artificially shifting the external trigger timing, a common stop signal of individual TDCs for each hit channel.  We found that hit timing should be corrected -0.7$\%$ to restore linearity.
                                                                                
\begin{figure}[t]                                                               
\centering                                                                      
\includegraphics[width=80mm]{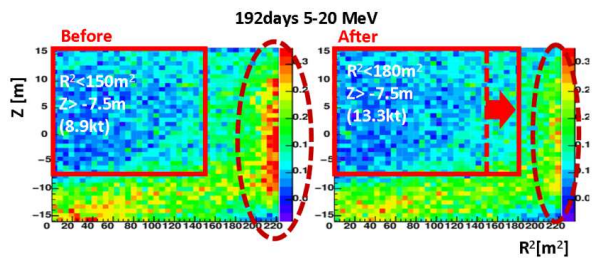}                                        
\caption{Vertex distribution before and after correction. }                     
\label{figver_vs}                                                               
\end{figure}

After the correction was applied the vertex shift shortened significantly (see Fig. \ref{figversh}).  As a result, the background events around the wall (inside brown ellipse) were reduced and the fiducial volume between 5.0 and 6.5 MeV could be enlarged up to 13.3 kton (Fig. \ref{figver_vs}). In addition, we can apply the same correction for SK-I/II and expect to reduce the systematic                    
uncertainty of the SK-I/II data, because the SK-III DAQ is the same as that of SK-I/II.     
           
\subsection{Time and Position Dependent MC\label{subsectpMC}}
\begin{figure}[b]
\centering
\includegraphics[width=80mm]{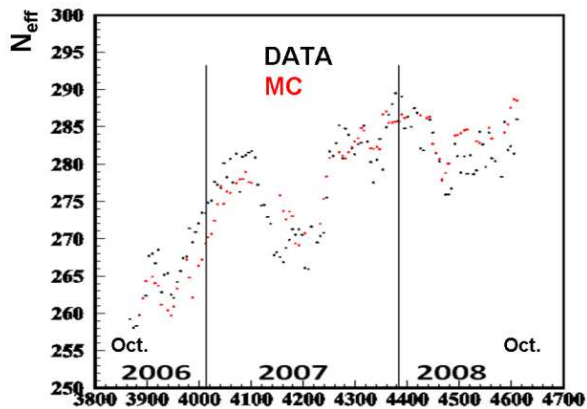}
\caption{$\mathrm{N_{eff}}$ time variation of Data (black) and MC (red) over the full SK-III period with fixed water transparency.} \label{figtimdecaye}
\end{figure}
\begin{figure}[t]
\centering
\includegraphics[width=80mm]{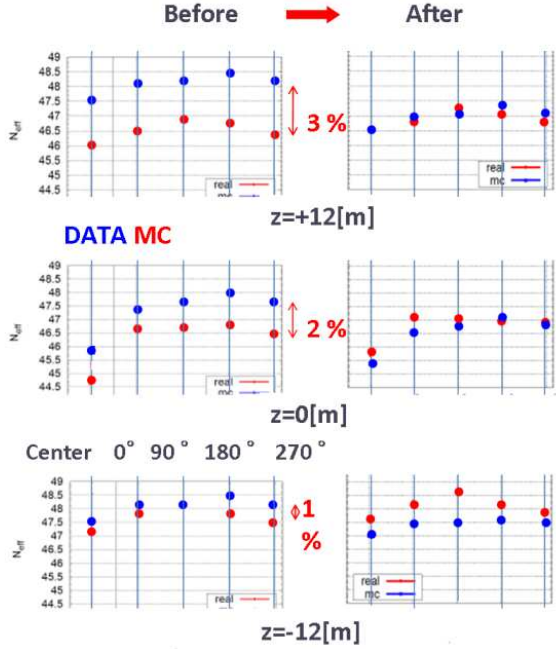}
\caption{$\mathrm{N_{eff}}$ of several positions before and after the position dependent MC installation}
\label{figdtp}
\end{figure}

That the water quality in SK changes as a function of time was known; the SK-I MC took it into account \cite{SKI_full}. In SK-III we also measured the time variation of the water transparency using $\mu$-$e$ decay data, as well as the wavelength dependence of various scattering and absorption water coefficients using a nitrogen/dye laser system \cite{SKnim}.

Using the full period of SK-III, we investigated possible relationships among the water transparency and water coefficients and found, similar to SK-I, the only apparent connection was between water transparency and the absorption coefficients. So we installed the absorption coefficients as a function of the water transparency measured by $\mu$-$e$ decay data and the scattering coefficients as a constant for the water transparency. As a result we had a time dependent MC.

We tested this MC with muon decay data. Figure  \ref{figtimdecaye} shows the $\mathrm{N_{eff}}$ time variation of data and MC in SK-III assuming a fixed water transparency. $\mathrm{N_{eff}}$ is the effective number of hits to yield the same value at any position in the SK tank by applying  several corrections to the number of hits \cite{SKI_full}.  One such correction takes into account the water transparency. To observe the effect of the time variation of water transparency directly, we didn't apply this correction, and confirmed that MC tracks the data properly.
 
We also found a position dependence in the water quality. The existence of position dependence due to the water flow has been debated since SK-I.  We installed light injectors on the barrel of the detector and tried to find it, but this method couldn't resolve the question.  Finally, in SK-III, using several calibration sources we concluded that position dependence not only exists but also  changes as a function of time. So we made a position dependent MC by installing a z-dependence in the water coefficients varied by the so-called top-bottom asymmetry as measured using calibration data over the full SK-III data-taking period. 

We tested this MC with \isotope[16]{N} calibration data \cite{SKI_full}. We took the data at several positions(Fig. \ref{figdtp}). Before correction the difference between data and MC was $\sim$3\% at the top of the detector, $\sim$2\% in the center, and $\sim$1\% at the bottom, but after correction was less than 1\% at the top and in the center. However, the difference at the bottom became larger; it still needs improvement.

\subsection{Angular Resolution\label{subsecar}}  
Another large source of systematic uncertainty in SK-I is the angular resolution \cite{SKI_full}. Because this uncertainty comes essentially from the hit pattern difference between data and MC, we must tune the MC to reduce the uncertainty. The MC was tuned several times. DAQ-related items like single photoelectron response, timing resolution, etc., and optical properties like scattering and absorption coefficients, the reflectance of PMTs, etc., were tuned using several calibration sources, as was done for SK-I \cite{SKnim}.                       
              
When we calculated the difference of the angular resolution of data and MC, their agreement was not improved compared with SK-I. So we explored several optical properties using MC, finally  finding that halving the original value for the reflectivity of a black sheet which covers the ID wall gave better agreement. Until that time we used the SK-II value for black sheet reflectivity, even though the materials of SK-II/III black sheet are different. 
                                            
\begin{figure}[t]                                                               
\centering                                                                      
\includegraphics[width=80mm]{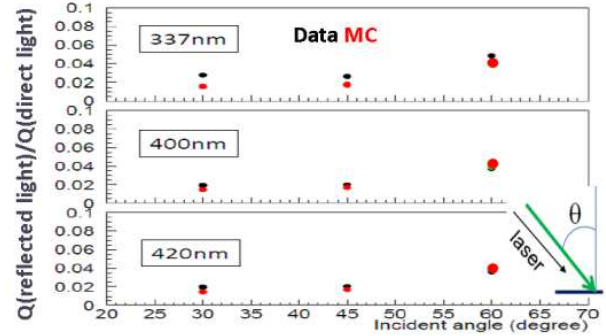}                                        
\caption{Black sheet reflectivity tuning result for three wavelengths. Data           
is black and MC is red.} \label{example_black}                                  
\end{figure}

To confirm this finding, we put a light injector with a black sheet reflector into the SK tank and measured the amount of direct light and reflected light for specific incident angles. Figure \ref{example_black} shows the ratio of the amount of charge due to reflected light versus direct light of data (black) and half reflectivity MC (red). This shows reasonable agreement, even if the  337 nm result (upper figure) still needs tuning.                                                
                                                                                
\begin{figure}[b]                                                               
\centering                                                                      
\includegraphics[width=80mm]{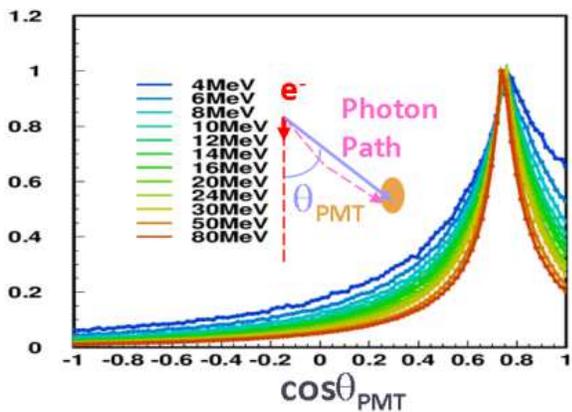}                                        
\caption{The distribution of the opening angle (depending on energy)              
between the particle direction and the vector from the reconstructed            
vertex to the hit PMT position.}                                                 
\label{example_like_dir}                                                        
\end{figure}                                                                    
                                                                                
\begin{figure}[t]                                                               
\centering                                                                      
\includegraphics[width=80mm]{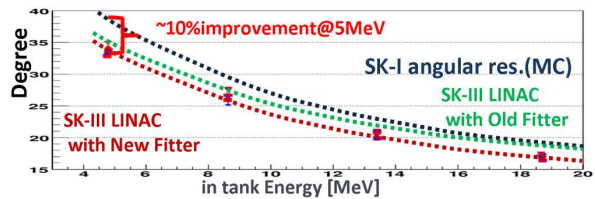}                                        
\caption{Angular resolution of SK-I/III.} \label{example_veres}                 
\end{figure}

Next, we improved the direction fitter itself. In essence, the likelihood function that represents the distribution of the opening angle between the particle direction and the vector from the reconstructed vetex to the hit PMT position was improved. In SK-I it was made for 10 MeV MC electrons \cite{SKI_full}, while for SK-III several energy bins were considered. As a result, we achieved about 10\% better angular resolution around 5 MeV than in SK-I  (Fig.  \ref{example_veres}).                     

\subsection{Other Items\label{subsecother}}
In addition to the above efforts, we retuned most of the reconstruction tools and reduction criteria to reduce their systematic uncertainties. We are continuing to tune those items whose tuning results weren't sufficiently good, as mentioned in the previous sections. As the determination of our  systematic errors is still under way, we can't show the exact numbers of the systematic uncertainties yet.

\section{New solar $\nu$ results\label{secres}}
As a result of the above efforts, we took good solar neutrino data during the full Super-Kamiokande-III data-taking period, which ran from August of 2006 through August of 2008. Currently 298 live days of data with a lower total energy threshold of 4.5 MeV is finished being analyzed. This section shows the following new solar neutrino results for this data set:
\begin{itemize}
\item Observed \isotope[8]{B} $\nu$ flux in SK-III
\item Angular distributions
\item \isotope[8]{B} $\nu$ flux time variation
\item Recoil electron energy spectrum
\item Day/Night asymmetry
\end{itemize}
But the 4.5-5.0 MeV energy bin still needs additional MC tuning, so for the flux calculation we didn't include this bin. We have only quoted statistical errors, since the estimation of the systematic uncertainties isn't finished.

\subsection{Observed the Solar $\nu$ Flux\label{subsecflux}}
\begin{figure}[t]
\centering
\includegraphics[width=80mm]{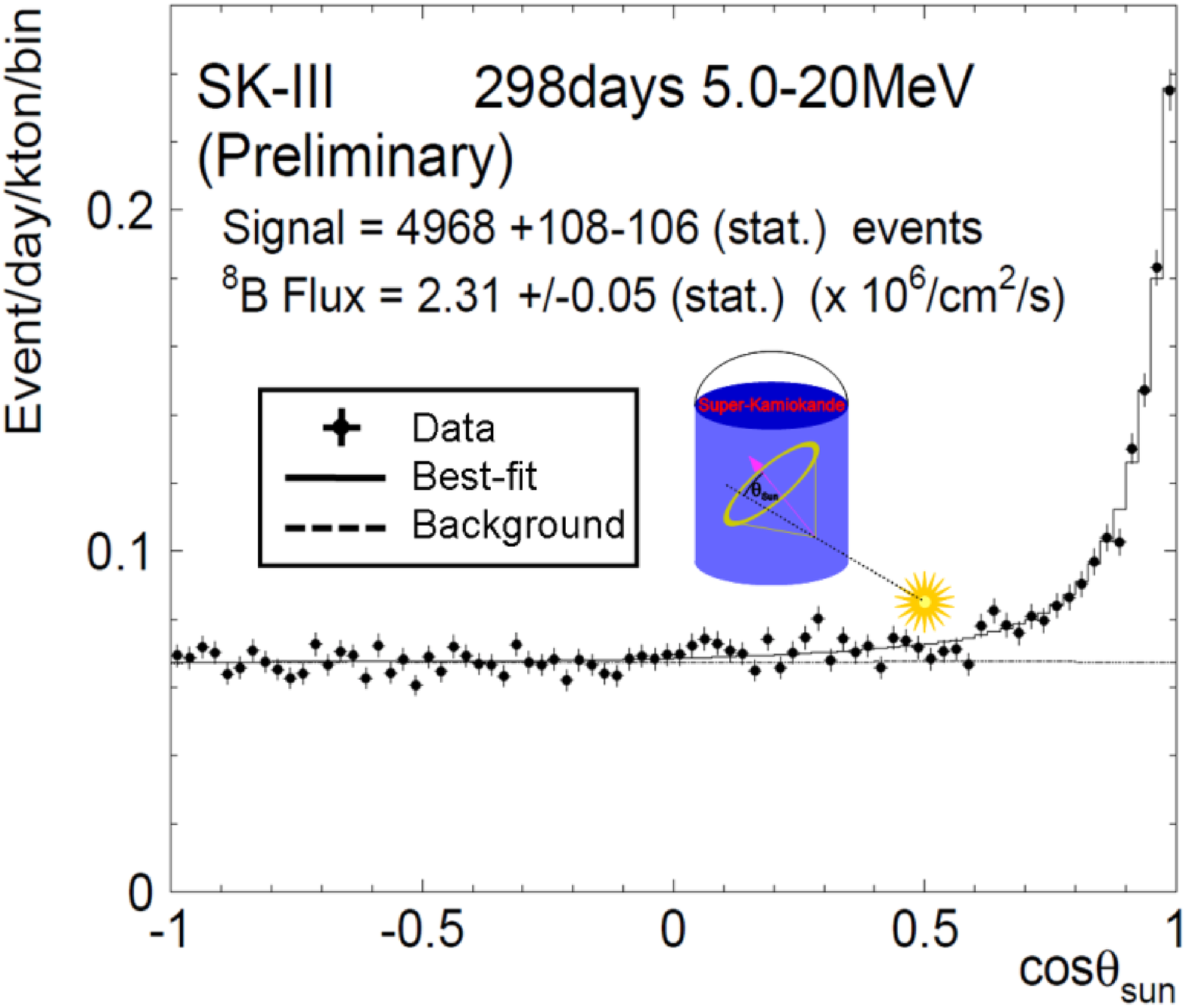}
\caption{Angular distribution of solar neutrino event candidates between 5.0 and 20.0 MeV.  The area under the dotted line is the contribution from remaining background events. The area between the solid and dotted line indicates the elastic scattering peak.} \label{example_figure}
\end{figure}
Solar neutrinos traversing SK occasionally interact with bound electrons of water molecules. These $\nu$-$e$ interactions are elastic scattering, and their incident neutrino and a recoil electron directions are highly correlated. Fig. \ref{example_figure} shows the $\cos\theta_{\mathrm{sun}}$ distribution of solar neutrino event candidates where  $\theta_{\mathrm{sun}}$ is the angle between the recoil electron direction and the direction from the sun; a sharp peak around 1.0 is observed. Using the same method as SK-I \cite{SKI_full}, we extracted the solar neutrino signal from this distribution. The best-fit value for the number of signal events between 5.0 MeV and 20.0 MeV  is $\mathrm{4968^{+108}_{-106}(stat.)}$. The corresponding \isotope[8]{B} $\nu$ flux is
\begin{equation*}
\mathrm{(2.31 \pm 0.05(stat.))\times 10^6cm^2s^{-1},}
\end{equation*}
is consistent with SK-I/II values of $\mathrm{(2.35 \pm 0.02(stat.) \pm 0.08(sys.)) \times 10^6cm^2s^{-1}}$ and $\mathrm{(2.38 \pm 0.05(stat.) ^{+0.16}_{-0.15}(sys.)) \times 10^6cm^2s^{-1}}$ within the statistical error.

\subsection{Angular Distributions\label{subsecang}}
Sec. \ref{subsecflux} showed the angular distribution of all events between 5.0 and 20.0 MeV. This section will focus on the angular distributions of the lowest energy bins.

\begin{figure}[t]
\centering
\includegraphics[width=80mm]{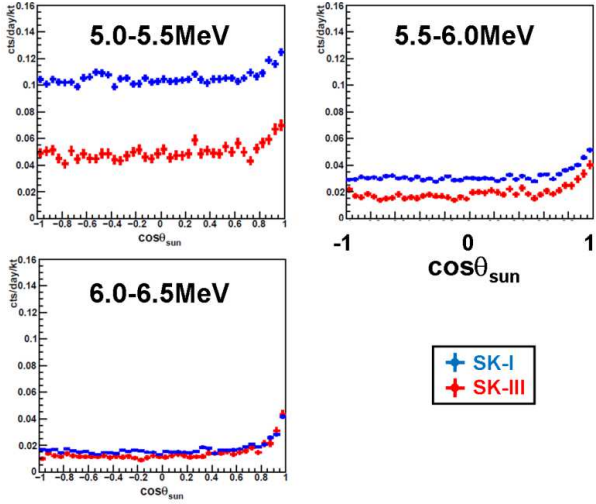}
\caption{SK-I(blue)/III(red) angular distributions of solar neutrino event candidates in 5.0-5.5, 5.5-6.0 and 6.0-6.5MeV regions and the same volume (the central 13.3 kton)} \label{example_figure_col2}
\end{figure}

 First, Fig. \ref{example_figure_col2} shows the 5.0-5.5, 5.5-6.0 and 6.0-6.5 MeV regions. For these energy bins, the fiducial volume (13.3 kton) is smaller than that (22.5 kton) of higher energy bins, because of the large number of background events from the ID wall (Sec. \ref{subsecWS}). As a result of the effort described in Sec. \ref{seccur}, the background level in SK-III became lower than that of SK-I, especially below 6.0 MeV where it has been cut by one half.  Moreover, the solar peak of SK-III is sharper than that of SK-I. In other words, the improved direction fitter (Sec. \ref{subsecar}) has made SK-III's angular resolution better than SK-I's.

\begin{figure}[t]
\centering
\includegraphics[width=80mm]{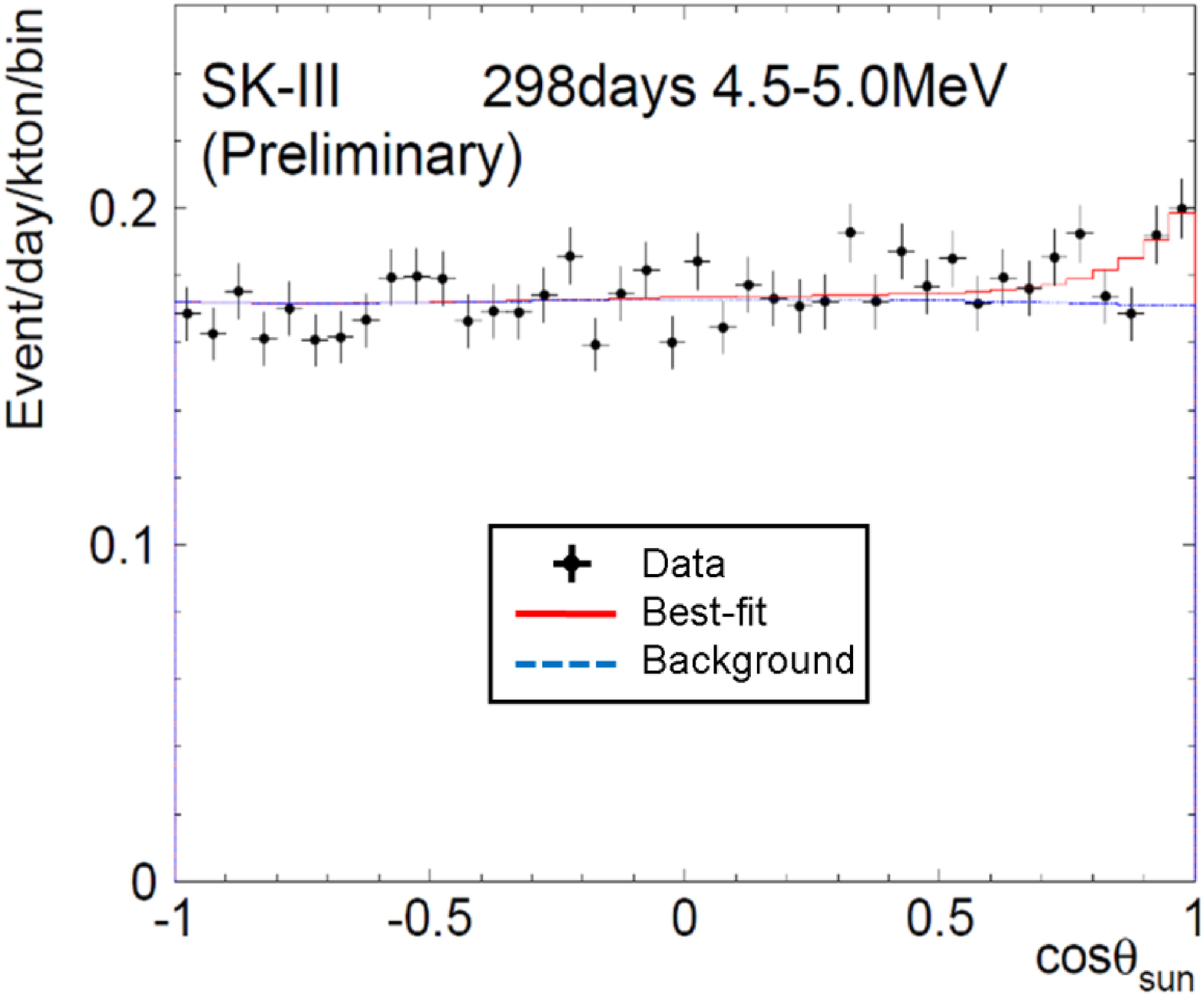}
\caption{Angular distribution of solar neutrino event candidates between 4.5 and 5.0 MeV. The area under the blue dotted line is the contribution from background events. The area between the red solid and blue dotted line indicates the elastic scattering peak.} \label{example_angle2}
\end{figure}

Next, we'd like to report a new low energy bin of 4.5-5.0 MeV. As a result of improvement of the SK water purification system, we were able to achieve the 4.5 MeV recoil electron energy threshold. Though high background from the ID wall make this fiducial volume (9.0 kton) much smaller than that of higher energy ranges, the low background in the central detector region provides a possibility to analyze the 4.5-5.0 MeV region; we have observed the solar signal in this energy bin. Figure \ref{example_angle2} shows the angular distribution of this region. Because of the huge background rate and insufficient exposure, at first glance the solar peak isn't readily apparent, but the fitting result was good. Still, due to the remaining work we cannot quote the exact flux number yet.

\subsection{Seasonal Variation\label{subsectime}}
\begin{figure}[t]
\centering
\includegraphics[width=80mm]{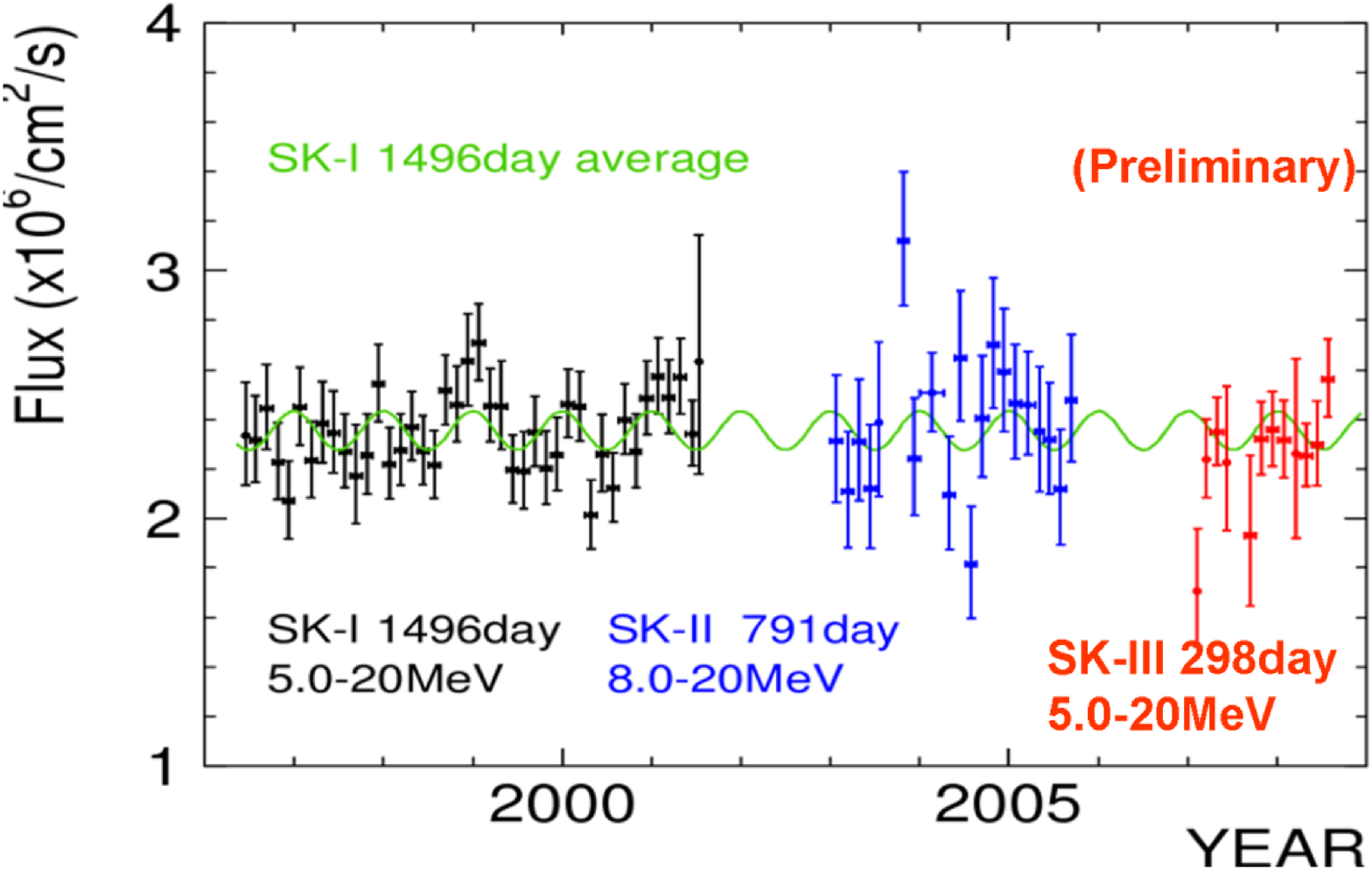}
\caption{The time variation of the solar flux beginning with SK-I.} \label{example_time}
\end{figure}

The time or seasonal variation of the total flux since the start of SK-I is shown in Fig. \ref{example_time}. The size of each horizontal bin is 1.5 months. The variation agrees well with a sinusoidal trend consistent with the expected $1/r^2$ variations in SK-I data due to the eccentricity of the Earth's orbit around the sun.

\subsection{Energy Spectrum\label{subsecspec}}
\begin{figure}[t]
\centering
\includegraphics[width=80mm]{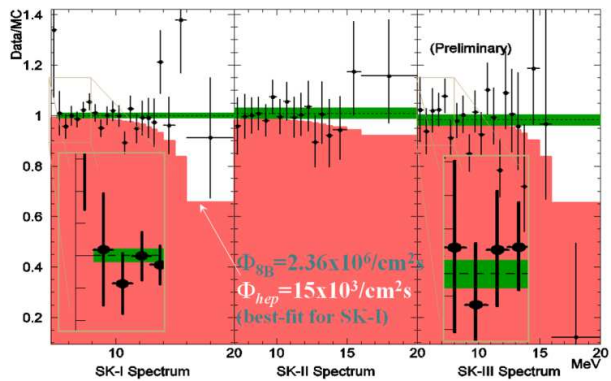}
\caption{Energy spectrum. Red area is \isotope[8]{B} $\nu$ contribution.} \label{example_spec}
\end{figure}
 Figure \ref{example_spec} shows the ratios of the expected (normalized by the SK-I best fit) MC and measured recoil electron energy spectra of SK-I/II/III. The dotted lines are the average data/MC ratios and the green bands represent statistical error. The red areas indicate the expected \isotope[8]{B} $\nu$ contribution from the SK-I best fit and the white areas under the dotted lines are the $hep$ $\nu$ contributions. We can confirm that SK-I/II/III are statistically consistent; even though SK-III has lower backgrounds below 6.5 MeV, statistical errors are bigger.

\subsection{Day/Night Asymmetry\label{subsecdna}}
 The Day/Night asymmetry($\mathrm{A_{DN}}$) is the only direct test of matter effects on the solar neutrino oscillations. This value is obtained from $\mathrm{A_{DN}=2(D-N)/(D+N)}$, where D and N are the day and night fluxes measured by selecting events which occur when the cosine of the solar zenith angle is less than zero (day) and greater than zero (night), respectively. In SK-I it was measured as $\mathrm{-2.1\% \pm 2.0\%(stat.)^{+1.3}_{-1.2}\%(syst.)}$, and also fitted as $\mathrm{-1.8\% \pm 1.6\%(stat.)^{+1.3}_{-1.2}\%(syst.)}$, and in SK-II measured as $\mathrm{-6.3\% \pm 4.2\%(stat.) \pm 3.7\%(syst.)}$. In SK-III it can be measured to 4.3\%(stat.) with the shown 298 days of data, and perhaps to 3.7\%(stat.) using the entire SK-III data set including periods of high threshold or high background runs. If so, then using the combined SK-I/II/III data can determine it to 1.6\%(stat.) and fit variations to 1.3\%(stat.).

\section{Conclusion\label{secsum}}
We achieved lower backgrounds below 6.0 MeV in the center of SK, and have almost finished analyzing SK-III data. We are trying to reduce systematic errors compared to SK-I and the estimation of these systematic errors is under way. Our SK-III results are consistent with the SK-I/II results within statistical uncertainties. At SK-III, the recoil electron energy threshold was lowered to 4.5 MeV. With existing data, SK is statistically sensitive down to LMA day/night asymmetries of 1.3\%. We expect to present preliminary SK-III solar neutrino results before the end of 2009.

\bigskip % extra skip inserted
% Create the reference section using BibTeX:
%\bibliography{basename of .bib file}

\begin{thebibliography}{9}   % Use for  1-9  references

\bibitem{SKI_full} Super-Kamiokande Collaboration: J. Hosaka et al., Phys. Rev. D 73, 112001 (2006).

\bibitem{SKII_full} Super-Kamiokande Collaboration: J.P. Cravens et al., Phys. Rev. D 78, 032002 (2008).

\bibitem{SNONCD} SNO Collaboration: B. Aharmim et al., Phys. Rev. Lett. 101, 111301 (2008).

\bibitem{KAMLAND} KamLAND collaboration: S. Abe et al., Phys. Rev. Lett. 100, 221803 (2008).

\bibitem{BOREXINO} Borexino Collaboration: arXiv:0808.2868.

\bibitem{nuclb143.13}  Super-Kamiokande Collaboration: M. Nakahata et al, Nucl. Phys. (Proc. Suppl.) B143, 13 (2005).

\bibitem{SKnim} Super-Kamiokande Collaboration: Y. Fukuda et al., Nucl. Instrum. Methods Phys. Res., Sect. A 501, 418 (2003).


%\begin{thebibliography}{99} % Use for 10-99 references

\end{thebibliography}

\end{document}